\documentclass[english,11pt,oneside]{article}
\pdfoutput=1
\usepackage{amsmath}
\usepackage{amsfonts}
\usepackage{amssymb}
\usepackage{graphicx, rotating}
\usepackage{epstopdf}
\usepackage{epsfig}
\usepackage{latexsym}
\usepackage{multirow}
\usepackage{color}
\usepackage{slashed}
\usepackage[top=2.8 cm, bottom=2.8 cm, left=3 cm, right=3 cm]{geometry}

\newcommand{\ba}{\begin{eqnarray}}
\newcommand{\ea}{\end{eqnarray}}
\newcommand{\no}{\nonumber}

\usepackage[colorlinks]{hyperref}
\hypersetup{
    citecolor={blue}
}

\begin{document}

\title{
The anomalous dimension of spin-1/2 baryons\\ in many flavors QCD
}
\author{Luca Vecchi$^{a,b,c}$\footnote{vecchi@pd.infn.it}\\
{\small\emph{$^a$ SISSA, via Bonomea 265, 34136, Trieste, Italy}}\\
{\small\emph{$^b$ Dipartimento di Fisica e Astronomia, Universit\`a di Padova and}}\\
{\small\emph{$^c$ INFN, Sezione di Padova, via Marzolo 8, I-35131 Padova, Italy
}}
}
\date{}
\maketitle

\begin{abstract}
The anomalous dimension of spin-1/2 baryon operators in QCD is derived at leading $1/N_f$ order using the minimal subtraction scheme. A residual ambiguity, originating from the presence of evanescent operators in dimensional regularization, is parametrized by a function of the renormalized coupling. Our result is shown to agree with previous 2 and 3 loop calculations performed in two different renormalization schemes. 
\end{abstract}

\newpage

\section{Introduction}
\label{intro}

Observables in QCD are functions of $\alpha=g^2/4\pi$ and $1/N_f$. An inspection of the 5-loop $\beta$ function~\cite{Baikov:2016tgj} (see also~\cite{Luthe:2016ima}), 5-loop $\gamma_m$~\cite{Baikov:2014qja}, and 3-loop $\gamma_\pm$~\cite{Gracey:2012gx} reveals that these RG functions may be re-organized in the ${\overline{\rm MS}}$ scheme as an expansion in $\alpha$ and $\sim N_f/10$ with coefficients of order unity or smaller. From this empirical observation we may conclude that ordinary perturbation theory should be reliable when $\alpha$ is sufficiently smaller than unity, whereas a large $N_f$ calculation should provide a reasonably accurate estimate of the exact, non-perturbative result for $N_f\gtrsim10$.

Nevertheless, investigations of QCD in the limit of large number $N_f$ of massless flavors are quite useful in practice. At the very least, they serve as non-trivial consistency checks for high-order calculations in ordinary perturbation theory. Furthermore, they represent an interesting laboratory for the study of dualities, see e.g.~\cite{Gracey:1996he} and references therein.

Yet, because even the first non-trivial order effectively re-sums an entire $\alpha N_f$ series, large $N_f$ calculations offer a unique, and systematically improvable probe of the non-perturbative regime of gauge theories. Therefore, despite the fact that for realistic numbers of massless flavors this approach is not fully justifiable,~\footnote{The authors of~\cite{Broadhurst:1994se} suggested the replacement $N_f\to-{3}\beta_0/2$, with $\beta_0=11-2N_f/3$ the coefficient of the QCD one-loop beta function, as a way to improve large $N_f$ computations to smaller $N_f$. This ``naive non-abelianization" effectively includes additional gluonic loops; however, it is an unsystematic (and gauge-dependent) truncation of the series and it is hard to judge its reliability. A more convincing way to quantify the impact of gluonic loops would be to calculate subleading $1/N_f$ corrections.} one can still hope that some quantity of physical interest be approximated reasonably well by the first few orders in $1/N_f$ even when $N_f\lesssim10$. Of course the QCD beta function cannot be such an example, since it changes abruptly with $N_c/N_f$. Other RG functions, such as the anomalous dimension of the quark bilinear or of baryons, might be more promising candidates.

An analysis of large $N_f$ QED was initiated in \cite{Espriu:1982pb}\cite{PalanquesMestre:1983zy} with the calculation of the anomalous dimension of the mass operator. The leading order result straightforwardly generalizes to large $N_f$ QCD. The beta function at next to leading order was calculated for QED in~\cite{PalanquesMestre:1983zy} and for QCD in~\cite{Gracey:1996he}. Here we wish to derive an intrinsically non-abelian quantity that has no counterpart in QED: the anomalous dimension of baryons. This is currently known up to 3 loops within standard perturbation theory~\cite{Gracey:2012gx}.

\section{Spin-1/2 operators in QCD}

We adopt a Weyl spinor notation, where all fermions are left handed: $\psi$ is a ${\bf3}$ of $SU(3)$ color and a fundamental of $SU(N_f)_L$ whereas $\widetilde{\psi}$ is a ${\overline{\bf3}}$ of $SU(3)$ color and the anti-fundamental of $SU(N_f)_R$. 

Using Fierz transformations it is easy to show that (in exactly $d=4$ dimensions) spin-1/2 baryons appear in two Lorentz structures:
\ba\label{basis}
[B_+]^{ijk}_\alpha=\psi^{\{i}_\alpha(\psi^{j\}}\psi^k)~~~~~~~~~~~~[B_-]^{i\tilde j\tilde k}_\alpha=\psi^i_\alpha(\widetilde{\psi}^{\tilde j}\widetilde{\psi}^{\tilde k})^*,
\ea
plus their conjugates. In this notation $(\psi_1\psi_2)=\psi_1^t\epsilon\psi_2$ --- with $\alpha, \beta, \cdots$ Lorentz indices and $\epsilon$ the fully antisymmetric 2 by 2 matrix ---, contractions of color is understood, and $i,j,k,\tilde i, \tilde j, \tilde k$ are flavor indices. The latter will often be suppressed for brevity, unless necessary to avoid ambiguities. We defined $\psi^{\{i}\psi^{j\}}\equiv\psi^i\psi^j+\psi^j\psi^i$.

The operators (\ref{basis}) are in different representations of the flavor group $SU(N_f)_L\times SU(N_f)_R$. $B_+$ transforms as the $\frac{1}{3}{{\bf N_f}({\bf N_f}^2-1)}$-dimensional representation of $SU(N_f)_L$, that has mixed symmetry properties, while $B_-$ is a $({\bf N_f},[{\bf N_f}\otimes {\bf N_f}]_{\rm antisym})$. 

In the chiral limit, and still in $d=4$ dimensions, $B_{\pm}$ do not mix under RG within any mass-independent renormalization scheme:
\ba\label{RG}
\mu\frac{d}{d\mu}
\left(\begin{array}{c}  
B^r_+ \\
B^r_-
\end{array}\right)
=-
\left(\begin{array}{cc}  
\gamma_+ &  \\
 & \gamma_-
\end{array}\right)
\left(\begin{array}{c}  
B^r_+ \\
B^r_-
\end{array}\right).
\ea
Here and in the following $B_\pm$ ($B_\pm^r$) denote the bare (renormalized) composite operators. By Parity conservation, similar relations hold for $\widetilde B_+=\widetilde{\psi}(\widetilde{\psi}\widetilde{\psi})$ and $\widetilde B_-=\widetilde{\psi}(\psi\psi)^*$, that have anomalous dimensions $\gamma_+,\gamma_-$ respectively.~\footnote{In the literature a different operator basis has often been adopted. A connection with the latter is straightforwardly obtained introducing a 4-component Dirac fermion $\Psi=(\Psi_L, \Psi_R)^t$ with $\Psi_L=\psi$, $\Psi_R=\epsilon\widetilde{\psi}^*$, and defining $B_1\equiv\Psi(\Psi^tC\Psi)$, $B_2=\gamma^5\Psi(\Psi^tC\gamma^5\Psi)$, where $C=i\gamma^0\gamma^2$. The latter basis is commonly used in lattice QCD simulations, since $B_{1,2}$ have the right Parity properties to interpolate a nucleon. ($B_1$ vanishes in the non-relativistic limit and therefore $B_2$ is usually preferred.) From the relations $(B_2\pm B_1)_L=-2B_\pm$, $(B_2\pm B_1)_R=+2\epsilon\widetilde B_\pm$, and (\ref{RG}), we see that:
\ba\label{constr}
\mu\frac{d}{d\mu}
\left(\begin{array}{c}  
B^r_1 \\
B^r_2
\end{array}\right)
=-\frac{1}{2}
\left(\begin{array}{cc}  
\gamma_++\gamma_- & \gamma_+-\gamma_- \\
\gamma_+-\gamma_- & \gamma_++\gamma_-
\end{array}\right)
\left(\begin{array}{c}  
B^r_1 \\
B^r_2
\end{array}\right).
\ea
Again, this expression is exact in the limit of unbroken chiral symmetry and Parity, $d=4$, and for any mass-independent scheme. The constraint (\ref{constr}) is consistently satisfied by the 3-loop calculation of~\cite{Gracey:2012gx}.}

Unfortunately, dimensional regularization (the mass-independent regularization scheme adopted here and virtually all multi-loop calculations), violates the assumption $d=4$. This introduces a mixing with evanescent operators with Lorentz structures such as $\Gamma\psi(\psi\Gamma\psi)$ and modifies (\ref{RG}) starting at 2-loops, as we discuss in detail next.

\subsection{Definition of the renormalization scheme}

Diagrams are regulated via dimensional regularization with $d=4-\varepsilon$ throughout the paper. Furthermore, we assume that the 2-dimensional matrices $\bar\sigma^\mu, \sigma^\mu$ and the anti-symmetric tensor $\epsilon$ are defined in $d$ dimensions. We use the notation of \cite{Dreiner:2008tw}.

Now, consider the following correlators of the bare operators:
\ba\label{short}
\langle B_+\rangle_{\dot\alpha\dot\beta\dot\gamma\delta}&\equiv&\langle{\psi^\dagger(p_1)}^i_{\dot\alpha a}{\psi^\dagger(p_2)}^{j}_{\dot\beta b}{\psi^\dagger(p_3)}^{k}_{\dot\gamma c}[B_+(-p_1-p_2-p_3)]^{i j k}_{\delta}\rangle\\\no
\langle B_-\rangle_{\dot\alpha\beta\gamma\delta}&\equiv&\langle{\psi^\dagger(p_1)}^i_{\dot\alpha a}{\widetilde\psi(p_2)}^{\tilde j}_{\beta b}{\widetilde\psi(p_3)}^{\tilde k}_{\gamma c}[B_-(-p_1-p_2-p_3)]^{i\tilde j\tilde k}_{\delta}\rangle,
\ea
where the repeated flavor indices are not summed. At leading order in $1/N_f$ we find:
\ba\label{loop}
\langle B\rangle^{(1)}&=&{\cal{D}}_{00}(\varepsilon,p)\langle B\rangle^{(0)}+{\cal{D}}_{01}(\varepsilon,p)\langle T\rangle^{(0)}.
\ea
In particular, the divergent part of $\langle B\rangle^{(1)}$ contains terms proportional to the tree correlator $\langle B\rangle^{(0)}\propto1\otimes1\otimes1$, as well as to non-trivial spinor structures like $\langle T\rangle^{(0)}\propto1\otimes\Gamma_{\mu\nu}\otimes\Gamma^{\mu\nu}+\Gamma_{\mu\nu}\otimes1\otimes\Gamma^{\mu\nu}+\Gamma_{\mu\nu}\otimes\Gamma^{\mu\nu}\otimes1$. The latter reduce to $1\otimes1\otimes1$ in $d=4$ dimensions, that is $T\to B$ as $\varepsilon\to0$. However, for $\varepsilon\neq0$ the Gamma matrices are not complete, $T$ is independent from $B$, and $B$ is not multiplicatively renormalized. In order to have a set of operators closed under RG we must extend (\ref{RG}) by introducing $T$, or more conveniently a linear combination $E_1$ of $T,B$ that vanishes as $\varepsilon\to0$, i.e. an evanescent operator. The latter would eventually mix with other evanescent operators involving a higher number of Gamma matrices and so on. The bottom line is that in dimensional regularization $B$ mixes with an infinite number of evanescent operators $E_{a=1,2,3,\cdots}$, invalidating (\ref{RG}). This complication is well appreciated in the context of 4-fermion operators, see e.g.~\cite{Bondi:1989nq}\cite{Dugan:1990df} for earlier literature, and~\cite{Herrlich:1994kh} for a lucid discussion.

Denoting the complete operator basis by $O_A$ ($=B,E_1,E_2,E_3,\cdots$), the bare and renormalized operators are related via 
\ba
O_A=Z_{AB}O^r_B, 
\ea
with $Z_{AB}=Z_{AB}^{\rm conn}Z_\psi^{3/2}$. By construction, the bare evanescent operators have vanishing tree-level matrix elements. On the other hand, the renormalized operators $E_a^r$ may contribute at loop level, though their matrix elements are not independent. In fact, in exactly 4-dimensions there exist finite functions $f_a$ of the renormalized coupling such that $\langle E^r_a\rangle=f_a\langle B^r\rangle$, see e.g.~\cite{Kraenkl:2011qb}. The functions $f_a$ are scheme-dependent. Fortunately, one can always choose a prescription where $f_a=0$.~\cite{Dugan:1990df} Such a scheme is especially useful when matching with a more fundamental theory at some high scale. The authors of~\cite{Dugan:1990df}\cite{Herrlich:1994kh} also found that $\gamma_{a0}=0$ in this case, so the running of the phenomenologically relevant parameters is simply controlled by the $00$ component of an infinite-dimensional anomalous dimension matrix $\gamma=Z^{-1}\mu dZ/d\mu$, i.e. $\gamma_{00}$, which itself receives contributions (starting at second nontrivial order) from loops involving evanescent operators. With this qualification (\ref{RG}) is correct. In a generic scheme with $f_a\neq0$ the evanescent operators also contribute to the matching. Moreover, the scaling of Green's functions with an insertion of $B^r$ is controlled by~\cite{Bondi:1989nq}\cite{Dugan:1990df} $\gamma=\gamma_{00}+\gamma_{0a}f_a$. Therefore (\ref{RG}) can also be made sense with $f_a\neq0$. 

None of this is relevant at leading order. Indeed, since $f_a$ first arises at ${\cal O}(1/N_f)$, we find $\gamma=\gamma_{00}+{\cal O}(1/N_f^2)$. Irrespective of $f_a$ we can thus write:
\ba\label{this}
\gamma^\pm
&=&\mu\frac{d}{d\mu}\left(\delta Z^{\rm conn}_{00\pm}+\frac{3}{2}\delta Z_\psi\right)+{\cal O}(1/N^2_f),
\ea 
where $\delta Z_{AB}=Z_{AB}-\delta_{AB}={\cal O}(1/N_f)$. At this order loops involving evanescent operators do not affect $\gamma_{00}$.

Yet, there is an additional, more subtle way in which the evanescent operators impact physical processes, which holds at any order and for any $f_a$. Indeed, the very definition of bare $E_a$ is not unique, and the choice we make ultimately affects the matrix elements of the renormalized physical operators $B_\pm^r$.~\cite{Herrlich:1994kh} In fact, in complete generality, we can {\emph{define}} 
\ba\label{bareE}
E_1=T-s(\varepsilon)B,
\ea
where $s(\varepsilon)=1+\sum_{n=1}s_n\varepsilon^n$ is an arbitrary function, and still satisfy the constraint $T\to B$ (as $\varepsilon\to0$). As a result, $B^r$ also depends on $s$ in general. Employing a minimal subtraction scheme, we go back to (\ref{loop}) and take:
\ba\label{Br}
\langle B_\pm^r\rangle={\rm finite}(1+{\cal{D}}_{00}+s_\pm(\varepsilon){\cal{D}}_{01})\langle B\rangle+{\cal O}(1/N_f).
\ea
Because in general both $B^r$ and $\gamma^\pm$ depend on $s$, a renormalization scheme is uniquely defined only once $s_\pm$ is given. This residual dependence on $s_\pm$ is discussed in section~\ref{sec:residual}.

\subsection{Anomalous dimension of the physical operators $B_\pm^r$}

Having introduced our renormalization prescription (\ref{Br}) we can now derive an explicit expression for (\ref{this}).

At leading $1/N_f$ order, the diagrams that contribute to $\gamma^\pm$ are the same as a 1-loop analysis, with the gluon propagator re-summing all fermion bubbles, see (\ref{AA}). Within dimensional regularization, and using the formulas (\ref{div}) and (\ref{div1}) from the Appendix, the divergent parts of the {\emph{connected}} diagrams contributing to $\langle B_\pm\rangle$ read (compare to (\ref{loop})):
\ba\label{tens}
{\rm div}\langle B_-\rangle^{\rm conn}_{\dot\alpha\beta\gamma\delta}&=&i(P_1)_{\alpha\dot\alpha}i(P_2)_{\beta\dot\beta}i(P_3)_{\gamma\dot\gamma}\epsilon_{abc}~{\rm div}\left[T^-_{\alpha\dot\beta\dot\gamma\delta\dot\sigma\dot\rho}\sum_{n=0}^\infty I_n+\xi~ {T'}^+_{\alpha\dot\beta\dot\gamma\delta\dot\sigma\dot\rho} I_0\right]\epsilon^{\dot\sigma\dot\rho}\\\no
{\rm div}\langle B_+\rangle^{\rm conn}_{\dot\alpha\dot\beta\dot\gamma\delta}&=&i(P_1)_{\alpha\dot\alpha}i(P_2)_{\beta\dot\beta}i(P_3)_{\gamma\dot\gamma}\epsilon_{abc}~{\rm div}\left[T^+_{\alpha\beta\gamma\delta\sigma\rho}\sum_{n=0}^\infty I_n+\xi~ {T'}^+_{\alpha\beta\gamma\delta\sigma\rho} I_0+(\rho\leftrightarrow\sigma)\right]\epsilon^{\sigma\rho},
\ea
where $P_i={\slashed{p_i}}/{p_i^2}$ is the tree-level fermion propagator, and $\xi$ the gauge parameter. The terms $\rho\leftrightarrow\sigma$ in the second line arise from the symmetrization of the $i j$ indices in the definition of $B_+$ (see (\ref{basis})). In the above expressions we introduced the tensorial structures
\ba\label{tens}
T^-_{\alpha\dot\beta\dot\gamma\delta\dot\sigma\dot\rho}&=&-\left[\delta_{\delta\alpha}(\bar\Gamma^{\mu\nu})_{\dot\beta\dot\sigma}(\bar\Gamma^{\mu\nu})_{\dot\gamma\dot\rho}-(\Gamma^{\mu\nu})_{\delta\alpha}\delta_{\dot\beta\dot\sigma}(\bar\Gamma^{\mu\nu})_{\dot\gamma\dot\rho}-(\Gamma^{\mu\nu})_{\delta\alpha}(\bar\Gamma^{\mu\nu})_{\dot\beta\dot\sigma}\delta_{\dot\gamma\dot\rho}\right]\\\no
{T}'^-_{\alpha\beta\gamma\delta\sigma\rho}&=&+3\left[\delta_{\delta\alpha}\delta_{\dot\beta\dot\sigma}\delta_{\dot\gamma\dot\rho}\right]\\\no
T^+_{\alpha\beta\gamma\delta\sigma\rho}&=&-\left[\delta_{\delta\alpha}(\Gamma^{\mu\nu})_{\sigma\beta}(\Gamma_{\mu\nu})_{\rho\gamma}+(\Gamma^{\mu\nu})_{\delta\alpha}\delta_{\sigma\beta}(\Gamma_{\mu\nu})_{\rho\gamma}+(\Gamma^{\mu\nu})_{\delta\alpha}(\Gamma_{\mu\nu})_{\sigma\beta}\delta_{\rho\gamma}\right]\\\no
{T}'^+_{\alpha\beta\gamma\delta\sigma\rho}&=&+3\left[\delta_{\delta\alpha}\delta_{\sigma\beta}\delta_{\rho\gamma}\right],
\ea
where the $d$ dimensional anti-symmetric tensors $\Gamma^{\mu\nu}, \bar\Gamma^{\mu\nu}$ are defined via $\sigma^\mu\bar\sigma^\nu=g^{\mu\nu}-2i\Gamma^{\mu\nu}$ and $\bar\sigma^\mu\sigma^\nu=g^{\mu\nu}-2i\bar\Gamma^{\mu\nu}$, whereas
\ba\label{In}
I_n&=&-\frac{2}{3}ig^2\frac{4}{d}\int\frac{{\rm d}^d\ell}{(2\pi)^d}\frac{\ell^4}{(\ell^2-\Delta)^4}\left[\Pi(\ell)\right]^n\\\no
&=&-\frac{1}{N_f}\left(-\frac{\lambda}{\varepsilon Z_A}\right)^{n+1}\frac{\overline\Pi^n}{n+1}\left(\frac{\mu^2}{\Delta}\right)^{(n+1)\frac{\varepsilon}{2}}
\frac{\left(1-\frac{\varepsilon}{6}\right)\Gamma\left(1+(n+1)\frac{\varepsilon}{2}\right)
}{\left(1+n\frac{\varepsilon}{2}\right)\left(1+n\frac{\varepsilon}{4}\right)\left(1+n\frac{\varepsilon}{6}\right)\Gamma\left(1+n\frac{\varepsilon}{2}\right)}.
\ea
with the factor of $2/3$ is due to the group theory identity $T^A_{a'a}[T^A_{bb'}]^*\epsilon^{a'b'c}=-\frac{2}{3}\epsilon^{abc}$, obtained for fermions in the fundamental representation ($T_R=1/2$). 

The quantity $\Delta$ depends on Lorentz-scalar combinations of the three momenta $p_{1,2,3}$ and therefore on the corresponding Feynman diagram. However, the results of Appendix~\ref{key} imply that $\Delta$ does appear in the divergent parts, as it must be in our regularization scheme. Because our main focus is the evaluation of the anomalous dimensions, $\Delta$ can therefore be ignored. It is still worth emphasizing that, as opposed to $\gamma_\pm$, the momentum-dependent finite terms are generically affected by renormalon poles~\cite{Broadhurst:1992si}.

As argued below (\ref{loop}), we are free to write $T^\pm$ as
\ba\label{mat}
T^\pm=3\left[s_\pm(\varepsilon)\right] 1\otimes1\otimes1 +T^\pm_{E}
\ea
for any non singular $s_\pm$ satisfying $s_\pm(0)=1$. Here $T^\pm$ are explicitly given in (\ref{tens}) whereas $T^\pm_{E}$ is the tensor structure associated to $E_1^\pm$. From our prescription (\ref{Br}) follows that $\delta Z_{00}={\rm div}({\cal{D}}_{00}+s{\cal{D}}_{01})$, where ${\cal{D}}_{00}+s{\cal{D}}_{01}$ in general depends on the external momenta, whereas $\delta Z_{00}$ does not. More explicitly:
\ba
\delta Z^{\rm conn}_{00,\pm}&=&{\rm div}\left[3s_\pm\sum_{n=0}^\infty I_n+3\xi I_0\right]+{\cal O}(1/N^2_f)\\\no
&=&-\sum_{n=1}^\infty\left(\frac{\lambda}{\varepsilon}\right)^n\frac{1}{n}G^{B}_0(\varepsilon)s_\pm(\varepsilon)+\frac{3\xi^r\lambda}{N_f}\frac{1}{\varepsilon}+{\cal O}(1/N_f^2),
\ea
with
\ba
G_0^{B}(\varepsilon)&=&-\frac{3}{N_f}\frac{\left(1-\varepsilon\right)\left(1-\frac{\varepsilon}{3}\right)\Gamma\left(1-\varepsilon\right)}{\left(1-\frac{\varepsilon}{2}\right)^2\left(1-\frac{\varepsilon}{4}\right)\Gamma\left(1+\frac{\varepsilon}{2}\right)\Gamma^3\left(1-\frac{\varepsilon}{2}\right)}.
\ea
In deriving $G_0^B$ we used the expression of $I_n$ given in (\ref{In}) and took advantage of (\ref{div}) (\ref{div1}).

Regarding the disconnected terms, note that the quark wave-function $Z_\psi$ can be calculated by replacing the tree gluon propagator with (\ref{AA}) in the familiar 1-loop diagram. As usual we define $Z_\psi=1+\delta Z_\psi=1+{\rm div}(\overline{\Sigma})+{\cal O}(1/N_f^2)$, where $\Sigma(q)\equiv\slashed{q}\overline{\Sigma}(q^2)$ is the 1-particle irreducible fermion 2-point function. Using the general formulas of Appendix~\ref{key} we find
\ba\label{psi}
Z_\psi&=&1-\frac{2\xi^r\lambda}{N_f}\frac{1}{\varepsilon}-\sum_{n=1}^\infty\left(\frac{\lambda}{\varepsilon}\right)^n\frac{1}{n}G^\psi_0(\varepsilon)+{\cal O}(1/N_f^2),\\\no
G_0^\psi(\varepsilon)&=&-\frac{3}{2N_f}\varepsilon\frac{\left(1-\varepsilon\right)\left(1-\frac{\varepsilon}{3}\right)^2\Gamma\left(1-\varepsilon\right)}{\left(1-\frac{\varepsilon}{2}\right)^2\left(1-\frac{\varepsilon}{4}\right)\Gamma\left(1+\frac{\varepsilon}{2}\right)\Gamma^3\left(1-\frac{\varepsilon}{2}\right)}.
\ea
This quantity was computed previously by other authors (see for instance~\cite{Gracey:1993ua} for a calculation in the $\xi=0$ gauge). Its determination does not present any subtlety associated to evanescent operators.

Plugging the above expressions for $G_0^{\psi,B}$ into (\ref{this}) and using (\ref{anom}) we arrive at our main result:
\ba\label{result}
\gamma_\pm(\lambda)&=&\lambda\left[G_0^{B}(\lambda)s_\pm(\lambda)+\frac{3}{2}G_0^\psi(\lambda)\right]\\\no
&=&-\frac{3}{N_f}\lambda\frac{\left(1-\lambda\right)\left(1-\frac{\lambda}{3}\right)^2\Gamma\left(1-\lambda\right)}{\left(1-\frac{\lambda}{2}\right)^2\left(1-\frac{\lambda}{4}\right)\Gamma\left(1+\frac{\lambda}{2}\right)\Gamma^3\left(1-\frac{\lambda}{2}\right)}\left(\frac{3}{4}\lambda+\frac{s_\pm(\lambda)}{1-\frac{\lambda}{3}}\right)+{\cal O}(1/N_f^2),
\ea
where $\lambda={\alpha_r N_f}/{3\pi}$ and $s_\pm(\lambda)=1+s_1^\pm\lambda+\cdots$. Consistently, $\gamma^\pm$ do not depend on $\xi^r$. This is because, in any mass-independent scheme (specifically the ${\overline{\rm MS}}$ scheme adopted here), the anomalous dimension of gauge invariant operators cannot depend on the gauge parameter. In our case, once this is verified at 1-loop, the result trivially extends to all terms at first order in $1/N_f$ because the longitudinal component of the gluon propagator is not renormalized, see (\ref{AA}).

\subsection{Two schemes for $s_\pm$}

We would like to compare (\ref{result}) to results obtained using standard perturbation theory. We consider the 2 and 3 loop calculations performed by~\cite{Kraenkl:2011qb} and~\cite{Gracey:2012gx}. We find that these are associated respectively to
\ba\label{f}
s_\pm=1~\cite{Kraenkl:2011qb},~\frac{d(d-1)}{12}~\cite{Gracey:2012gx}.
\ea

The fact that $s_\pm=1$ in~\cite{Kraenkl:2011qb} follows immediately from~(\ref{tens}) and the subtraction scheme introduced in that reference. Moreover, using (\ref{tens}) we see that the anomalous dimension of the general operator $O=\psi_\alpha\psi_\beta\psi_\gamma$ in that scheme is:
\ba\label{resultp}
\gamma_O&=&-\frac{3}{N_f}\lambda\frac{\left(1-\lambda\right)\left(1-\frac{\lambda}{3}\right)^2\Gamma\left(1-\lambda\right)}{\left(1-\frac{\lambda}{2}\right)^2\left(1-\frac{\lambda}{4}\right)\Gamma\left(1+\frac{\lambda}{2}\right)\Gamma^3\left(1-\frac{\lambda}{2}\right)}\left(\frac{3}{4}\lambda~{\mathbb{C}_0}+\frac{1}{3(1-\frac{\lambda}{3})}~{\mathbb{C}_2}\right)\\\no
&+&{\cal O}(1/N_f^2),
\ea
where 
\ba\no
{\mathbb{C}_0}&=&1\otimes1\otimes1\\\no
{\mathbb{C}_2}&=&1\otimes\Gamma_{\mu\nu}\otimes\Gamma^{\mu\nu}+\Gamma_{\mu\nu}\otimes1\otimes\Gamma^{\mu\nu}+\Gamma_{\mu\nu}\otimes\Gamma^{\mu\nu}\otimes1. 
\ea
The 4-dimensional scalings of the 3-quark operators of spin $1/2, 3/2$ are obtained replacing ${\mathbb{C}_2}=+3,-3$, respectively, and ${\mathbb{C}_0}=1$ in both of them (there is a factor of $1/2$ difference in our definition of $\Gamma^{\mu\nu}$ compared to~\cite{Kraenkl:2011qb}). An analogous expression holds for $\widetilde{O}=\psi_\alpha({\widetilde{\psi}}_{\dot\beta}{\widetilde{\psi}}_{\dot\gamma})^*$.

To see why $s_\pm={d(d-1)}/{12}$ characterizes the formalism of~\cite{Gracey:2012gx} is a bit more complicated. Rather than repeating the calculation using the operator basis defined there, we can get to~(\ref{f}) observing that in the basis used by ref.~\cite{Gracey:2012gx} only the spinor structure $1\otimes1\otimes1$ appears at order $1/N_f$, so the expression corresponding to (\ref{mat}) simply reads $T^\pm_{\rm ref}=3\left[s_\pm^{\rm ref}(\varepsilon)\right] 1\otimes1\otimes1$. Obviously, if we contract the Lorentz indices in $\langle B_\pm\rangle$ with the external momenta as 
\ba\label{contraction1}
p_1^\mu\bar{\sigma}_\mu^{\dot\alpha\delta}\epsilon^{\dot\beta\dot\gamma} \langle B_+\rangle_{\dot\alpha\dot\beta\dot\gamma\delta},~~~~~~~~~p_1^\mu\bar{\sigma}_\mu^{\dot\alpha\delta}\epsilon^{\beta\gamma} \langle B_-\rangle_{\dot\alpha\beta\gamma\delta},
\ea
the resulting expressions are multiplicatively renormalized as well. What is less obvious is that also within our formalism the contractions (\ref{contraction1}) are multiplicatively renormalized. This follows from the fact that traces of the Gamma matrices can be simplified in any dimension using $\left\{\sigma^\mu,\bar\sigma^\nu\right\}=2g^{\mu\nu}$. (As a matter of fact, all tensor structures in (\ref{tens}) become products of identities when contracted with ${\slashed{\bar p_1}}\otimes\epsilon$). These contractions, being Lorentz scalar combinations of the external momenta, do not depend on our basis of fermionic operators. Hence, making the identification ${contraction}(T^\pm)={contraction}(T^\pm_{\rm ref})$ we obtain $s_\pm^{\rm ref}=d(d-1)/12$, as anticipated. That this factor appears in the comparison between the two schemes was already stressed in~\cite{Gracey:2012gx}.

\begin{figure}[t]
\begin{center}
\includegraphics[width=9.5cm]{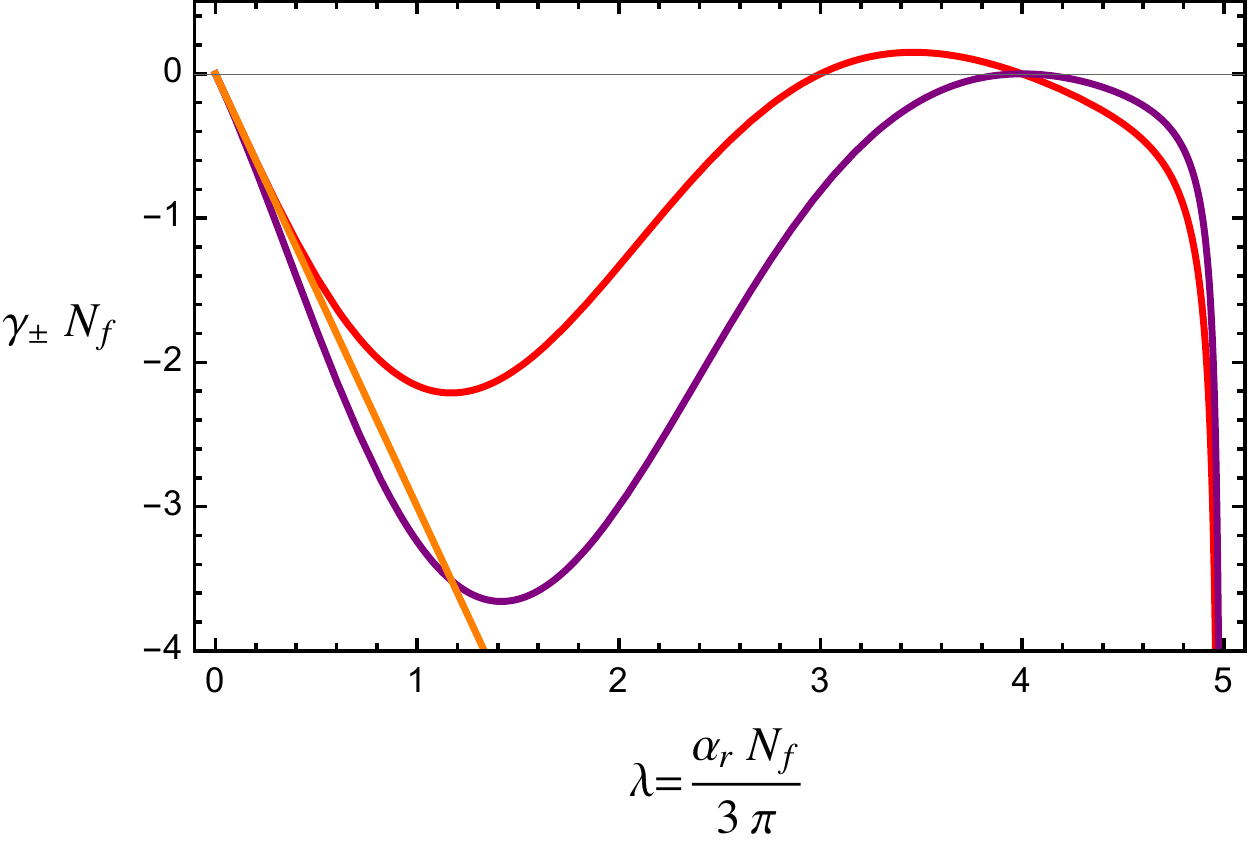}
\caption{Anomalous dimension of baryons in the ${\overline{\rm MS}}$ scheme at leading order in $1/N_f$ and all orders in the coupling $\lambda=\alpha_r N_f/3\pi$ for different definitions of evanescent operators, $s_\pm=1$ (purple)~\cite{Kraenkl:2011qb}, $s_\pm=d(d-1)/12$ (red)~\cite{Gracey:2012gx}, and $s_\pm$ such that $\gamma^\pm$ equals the 1-loop result (orange).
}\label{plot}
\end{center}
\end{figure}

Expanding $\gamma^\pm$ in powers of $a=\alpha_r/4\pi$ we get:
\ba
\gamma_\pm^{\left(s_\pm=\frac{d(d-1)}{12}\right)}&=&-4 a - \frac{4}{9}N_f a^2 + \frac{260}{81} N_f^2 a^3+\frac{4}{81} \left(51- 48\,\zeta_3\right)N_f^3a^4+{\cal O}(N_f^4a^5),\\\no
\gamma_\pm^{\left(s_\pm=1\right)}&=&-4 a - \frac{32}{9}N_f a^2 + \frac{112}{27} N_f^2 a^3+\frac{64}{81} \left(5- 3\,\zeta_3\right)N_f^3a^4+{\cal O}(N_f^4a^5),
\ea
with $\zeta_3=1.20206\cdots$. The first terms agree with the calculation of~\cite{Kraenkl:2011qb} and~\cite{Gracey:2012gx}.~\footnote{For the latter, this is true up to an overall factor of $-2$ arising from a different definition of $\gamma_\pm$. Our conventions conform with those adopted in the one-loop analysis of~\cite{Peskin:1979mn} and the two-loop calculation of~\cite{Pivovarov:1991nk} ($N_f=3$) and~\cite{Aoki:2006ib} (general $N_f$), as well as~\cite{Kraenkl:2011qb}.} This provides a non-trivial check of our result. The full expression (\ref{result}) is shown in figure~\ref{plot}. For generic $s_\pm$ it has a simple pole at $\lambda=5$, that also sets the radius of convergence of the $\lambda$ series.

\subsection{Independence of observables on $s_\pm$}
\label{sec:residual}

The scheme-dependent function $s_\pm$ appearing in the definition of bare evanescent operators affects the matrix elements of the renormalized physical operator, see (\ref{Br}), as well as its running, see (\ref{result}). Specifically, from (\ref{Br}) and recalling that $s=1+\sum_{n=1}^\infty s_n\varepsilon^n$,
\ba\label{Bra}
\frac{d}{ds_n}\ln\langle B^r\rangle
&=&\frac{d}{ds_n}{\rm finite}(s{\cal{D}}_{01})+{\cal O}(1/N_f)\\\no
&=&-\sum_{k=0}^\infty[G_0^B]_k\frac{\lambda^{n+k}}{n+k}+{\cal O}(1/N_f),
\ea
with $G_0^B(\varepsilon)=\sum_{k=0} [G_0^B]_k\varepsilon^k$. Because only the {\emph{divergent}} part of ${\cal{D}}_{01}$ contributes to this expression, (\ref{div1}) can be employed to show that $\frac{d}{ds_n}\ln\langle B^r\rangle$ is momentum independent. A more convenient way to write the above law is obtained observing that 
\ba\label{thanks}
\sum_{k=0}^\infty[G_0^B]_k\frac{\lambda^{n+k}}{n+k}=\int^{\lambda}_{0} d\lambda~G_0^B(\lambda)\lambda^{n-1}=\frac{d}{ds_n}\int^{\lambda}_{0} d\lambda\frac{\gamma}{\lambda^2}.
\ea

Of course physical quantities do not depend on our renormalization scheme, so the dependence of $\langle B^r\rangle$ on $s_\pm$ must cancel that in corresponding Wilson coefficients. To show how this cancellation works to all orders in $\alpha N_f$, consider an effective field theory defined by the following effective ($d=4-\varepsilon$ dimensional) Hamiltonian:
\ba
{\cal H}=C^rB^rJ+\sum_{a=1} C^r_aE^r_aJ_a. 
\ea
Here $J,J_a$ are fermionic currents that excite our baryonic operators. We are interested in processes with a single insertion of ${\cal H}$. Within the prescription of~\cite{Dugan:1990df}, where $f_a=0$, $C^r$ is the only physical Wilson coefficient and is derived matching $\langle{\cal H}\rangle=C^r\langle B^rJ\rangle$ onto some physical process. The RG improved coefficient reads
\ba\label{CRG}
C^r(\mu, s)=e^{\int^{\lambda(\mu)}_{\lambda(\mu')} d\lambda\frac{\gamma}{\lambda^2}}C^r(\mu',s),
\ea
where we used $\mu d\lambda/d\mu=\lambda^2$. By RG invariance of $\langle{\cal H}\rangle$ we know that $\frac{d}{ds_n}\ln C^r(\mu, s)$ depends on $\mu$ only via $\lambda(\mu)$. An inspection of~(\ref{CRG}) reveals that this latter constraint holds as long as
\ba\label{schemeDep}
\frac{d}{ds_n}\ln C^r(\mu, s)=\frac{d}{ds_n}\int^{\lambda}_{0} d\lambda\frac{\gamma}{\lambda^2}.
\ea
Thanks to (\ref{Bra}) and (\ref{thanks}) this is equivalent to $\frac{d}{ds_n}\langle {\cal H}\rangle=0$, as expected. Note that the right hand side of (\ref{Bra}) and (\ref{schemeDep}) are ${\cal O}(\lambda)$ because the scheme-dependence of $\gamma$ starts at 2-loops.

We end with a comment on the conformal window of many flavors QCD. It is well known that the QCD beta function has zeros at $N_f^c\leq N_f\leq16$, for some unknown number $N_f^c$. At these IR fixed points, critical exponents like $\gamma^\pm$ become physical, scheme-independent quantities. However, the scheme-independence of (\ref{result}) might seem surprising given that in $\overline{\rm MS}$ the renormalized coupling does not carry any information about the evanescent operators. This puzzle is solved observing that the defining condition $\beta(\lambda_*)=0$ requires cancellations between terms of different order in the $1/N_f$ expansion; it then becomes possible for terms of different order in $1/N_f$ to conspire so as to remove any $s_n$-dependence from $\gamma^\pm(\lambda_*)$. From this observation we learn two things. First, the next to leading terms in the large $N_f$ expansion of $\gamma^\pm$ must also depend on the $s_n$'s of~(\ref{result}). This is necessary for the above cancellation to take place. Second, the variation in $\gamma^\pm(\lambda_*)$ as a function of $s_\pm$, see figure~\ref{plot}, should give us a rough estimate of the size of the next to leading corrections within the conformal window.

\section{Discussion}

The anomalous dimension of the QCD spin-1/2 baryons, at ${\cal O}(1/N_f)$ and all orders in $\lambda=\alpha_r N_f/3\pi<5$, can be written in the minimal subtraction scheme as:
\ba
\gamma_\pm(\lambda)=\frac{1}{2}\gamma_m(\lambda)\left(\frac{3}{4}\lambda+\frac{s_\pm(\lambda)}{1-\frac{\lambda}{3}}\right)+{\cal O}(1/N_f^2),
\ea
where $\gamma_m$ is the anomalous dimension of the mass operator, first calculated in \cite{PalanquesMestre:1983zy}. (In our notation the scaling dimension of the quark bilinear is $3+\gamma_m$ while that of baryons $4.5+\gamma_\pm$.) 

Here $s_\pm(\lambda)$ are scheme-dependent functions of the renormalized coupling satisfying $s_\pm(0)=1$. The residual scheme-dependence it entails stems from an ambiguity in the definition of the evanescent operators introduced in dimensional regularization. The two schemes adopted in~\cite{Kraenkl:2011qb}\cite{Gracey:2012gx} correspond to $s_\pm=1, (1-\lambda/3)(1-\lambda/4)$, respectively. 

The functions $s_\pm$ also affect the matrix elements of the renormalized physical operators $B_\pm^r$. However, in any observable such dependence is exactly compensated by that of $\gamma^\pm$. We explicitly saw how this works to all orders in $\alpha_rN_f$.

The anomalous dimensions $\gamma^\pm$ have a phenomenological application in scenarios beyond the Standard Model, for example in the calculation of the proton decay rate~\cite{Abbott:1980zj}. They are also relevant in scenarios with exotic QCD-like dynamics in the conformal window~\cite{Vecchi:2015fma}. Unfortunately, compared to a 1-loop estimate, Eq.~(\ref{result}) does not provide any quantitatively trustable information about their actual size because of the intrinsic non-pertubative nature of the conformal window. Indeed, for $13\leq N_f\leq16$ perturbation theory is reliable and even the scheme-independent one-loop result $\gamma^\pm=-\alpha/\pi$ is accurate. For example, at the zero of the 5-loop beta function with $N_f=13$, $\alpha_*=0.406$, the values of $\gamma^\pm(\lambda_*)$ lie within $\gamma_*=-(0.12\div0.15)$, consistently with \cite{Kraenkl:2011qb}\cite{Gracey:2012gx}. When $N_f<13$ ordinary perturbation theory becomes unreliable, as testified by the fact that the IR fixed point found at 2,3,4-loops disappears at 5-loops. Similarly, the residual scheme-dependence in (\ref{result}), argued to be of order $3/N_f$ in the previous section, quickly becomes uncomfortably large.

\section*{Acknowledgments}

We thank J. Gracey for comments and suggestions, as well as G. Ferretti and A. G. Grozin for discussions. We acknowledge the Mainz Institute for Theoretical Physics (MITP) for its kind hospitality and support during the final stages of this work. LV is supported by the ERC Advanced Grant no.267985 ({\emph{DaMeSyFla}}) and the MIUR-FIRB grant RBFR12H1MW.

\appendix

\section{Large $N_f$ at leading order}
\label{key}

At leading order in $1/N_f$ QCD correlators have a very simple structure. The leading diagrams may be simply derived by replacing the bare gluon propagator in a 1-loop analysis with an improved bare quantity:
\ba\label{AA}
\langle A_\mu A_\nu\rangle&=&-\frac{i}{q^2}\left(g_{\mu\nu}-\frac{q_\mu q_\nu}{q^2}\right)\sum_{n=0}^\infty\Pi^n-\frac{i}{q^2} \xi \frac{q_\mu q_\nu}{q^2},
\ea
where
\ba
\Pi&=&-\frac{\lambda}{\varepsilon Z_A}\left(-\frac{\mu^2}{q^2}\right)^{\varepsilon/2}\overline\Pi(\varepsilon),\\\no
\overline\Pi(\varepsilon)&\equiv&
\frac{\left(1-\frac{\varepsilon}{2}\right)^2\Gamma\left(1+\frac{\varepsilon}{2}\right)\Gamma^2\left(1-\frac{\varepsilon}{2}\right)}{\left(1-\varepsilon\right)\left(1-\frac{\varepsilon}{2}\right)\left(1-\frac{\varepsilon}{3}\right)\Gamma\left(1-\varepsilon\right)}.
\ea
In the previous expression we introduced the renormalized coupling and the gluon wave-function
\ba\label{xZ}
\lambda=T_R\frac{2\alpha_r N_f}{3\pi},~~~~~~~~~~~~~~Z_A=1-\frac{\lambda}{\varepsilon}
\ea
with ${\rm tr}(T^aT^b)=T_R\delta^{ab}$ for a given fermion color representation with generators $T^a$ ($T_R=1/2$ for the fundamental representation, whereas $T_R=1$ in QED). The renormalized coupling $\alpha_r=\mu^{-\varepsilon}Z_Ag^2(4\pi)^{-1+\varepsilon/2}$ ($g$ is the bare coupling) satisfies
\ba\label{beta}
\mu\frac{d\lambda}{d\mu}=-\varepsilon \lambda Z_A=(\lambda-\varepsilon)\lambda.
\ea
Eq.(\ref{AA}) includes all fermion bubbles and therefore re-sums the entire $\lambda$ series. At the order we are working the relation between the bare and renormalized vector fields is $A_\mu=\sqrt{Z_A}A_\mu^r$, whereas the gauge parameter is renormalized according to $\xi=Z_A\xi^r$.

All $1/N_f$ diagrams are found to be of the form: 
\ba\label{div}
C=\sum_{n=1}^\infty\left(\frac{\lambda}{\varepsilon Z_A}\right)^n\frac{(-1)^n}{n}G(\varepsilon,n\varepsilon).
\ea
This key relation, first appeared in the calculation of the anomalous dimension of the fermion mass operator~\cite{PalanquesMestre:1983zy}, allows us to systematically analyze diagrams at the leading $1/N_f$ order.

Once our diagrams are put in the standard form (\ref{div}), and $G(\varepsilon,n\varepsilon)=\sum_{k=0}^\infty G_k(\varepsilon)(n\varepsilon)^k$ is understood as a series (we motivate this assumption below), the extraction of the anomalous dimension becomes a trivial task. Indeed, expanding in powers of the renormalized coupling (recall (\ref{xZ})) using the negative binomial series we find that the divergent part in perturbation theory is just given by:
\ba\label{div1}
{\rm div}(C)=-\sum_{n=1}^\infty\left(\frac{\lambda}{\varepsilon}\right)^n\frac{1}{n}G_0(\varepsilon).
\ea
If we now write $G_0(\varepsilon)=\sum_{n=0}^\infty g_n\varepsilon^n$ as yet another formal series (again to be motivated shortly), and employ (\ref{beta}), we finally arrive at a very useful compact expression:
\ba\label{anom}
\mu\frac{d}{d\mu}{\rm div}(C)=\lambda\sum_{n=0}^\infty g_n\lambda^n=\lambda\,G_0\left(\lambda\right).
\ea
Eq. (\ref{anom}) is just a manifestation of the fact that it is the coefficient of $1/\varepsilon$ in the correlator $C$ that contains information on the RG evolution~\cite{'tHooft:1973mm}. 

We can now motivate our expansion in $\varepsilon$. Eqs. (\ref{div}) and (\ref{anom}) show that the dependence on $\varepsilon$ of the regulated Feynman diagrams secretly encode the perturbative series in $\lambda$. We may interpret this as an indication that the very existence of ordinary perturbation theory justifies our series expansion in $\varepsilon$. Physically, the relation between the $\varepsilon$ and $\lambda$ expansions stems from the fact that in $d=4-\varepsilon$ dimensions, large $N_f$ QCD has a non-trivial IR fixed point at $\lambda^*=\varepsilon+{\cal O}(1/N_f)$, see (\ref{beta}): because anomalous dimensions of gauge-invariant operators are functions solely of the renormalized coupling, the critical exponents $\gamma(\varepsilon)$ in $d=4-\varepsilon$ dimensions can be trivially mapped onto $\gamma(\lambda)$, up to subleading $1/N_f$ corrections. This approach has been applied to large $N_f$ QCD in~\cite{Gracey:1991xf}.

 \end{document}